\documentclass[twocolumn,pre,showpacs,nofootinbib]{revtex4}
\usepackage{dcolumn}
\usepackage{amsmath}
\usepackage{amssymb}
\usepackage{calrsfs}
\usepackage{graphicx}
\usepackage[linkcolor=red,anchorcolor=blue,citecolor=green, bookmarks=false]{hyperref}

\newcommand{\lar}{\overleftarrow}
\newcommand{\rar}{\overrightarrow}

\begin{document}

\title{Stochastic block model and exploratory analysis in signed networks}

\author{Jonathan Q. Jiang}
\altaffiliation[Current address: ]
{Department of Mathematics, Hong Kong Baptist University, Kowloon Tong, Hong Kong}
\affiliation{Department of Computer Science, City University of Hong Kong, 83 Tat Chee Avenue,
Kowloon, Hong Kong}

\begin{abstract}
We propose a generalized stochastic block model to explore the mesoscopic structures in signed networks by grouping vertices that exhibit similar positive and negative connection profiles into the same cluster. In this model, the group memberships are viewed as hidden or unobserved quantities, and the connection patterns between groups are explicitly characterized by two block matrices, one for positive links and the other for negative links. By fitting the model to the observed network, we can not only extract various structural patterns existing in the network without prior knowledge, but also recognize what specific structures we obtained. Furthermore, the model parameters provide vital clues about the probabilities that each vertex belongs to different groups and the centrality of each vertex in its corresponding group. This information sheds light on the discovery of the networks¡¯ overlapping structures and the identification of two types of important vertices, which serve as the cores of each group and the bridges between different groups, respectively. Experiments on a series of synthetic and real-life networks show the effectiveness as well as the superiority of our model.
\end{abstract}


\maketitle

\section{Introduction}
\label{sec1}

The study of networks has received considerable attention in recent literature~\cite{Newman03,MJP09,Fortunato10}. This is mainly attributed to the fact that a network provides a concise mathematical representation for social~\cite{GAT07,AEPM08}, technological~\cite{GSCF02}, biological~\cite{RL05,GIIT05,MP07} and other complex systems~\cite{Newman03,MJP09,Fortunato10} in the real world, which paves the way for executing proper analysis of such systems' organizations,
functions and dynamics.

Many networks are found to possess a multitude of mesoscopic structural patterns, which can be coarsely divided into \lq\lq assortative\rq\rq~or \lq\lq community\rq\rq~structure and \lq\lq
disassortative\rq\rq~or \lq\lq bipartitie/multipartite\rq\rq~structure~\cite{Newman06,ME07}. In addition, other types of mesoscopic structures, such as the \lq\lq core-periphery\rq\rq~motif,
have been observed in real-life networks as well. Along with these discoveries, a large number of techniques have been proposed for mesoscopic structure extraction, in particular for community detection (see, e.g.~\cite{GIIT05,Newman06,ME07,MM02,PFCB03,MM04} and recent
reviews~\cite{Fortunato10,MJP09,LJAA05}). Most, if not all, existing techniques require us to know which specific structure we are looking for before we study it. Unfortunately, we often know little about a given network and have no idea what specific structures can be expected and subsequently detected by what specific methods. Biased results will be obtained if an inappropriate method is chosen. Even if we know something beforehand, it is still difficult for a method that is exclusively designed for a certain type of mesoscopic structure to uncover the aforementioned miscellaneous structures that may simultaneously coexist in a network or may even overlap with each other~\cite{GIIT05,DJP05,TAF08,ISMB11,CSA11,JL12}.

To overcome these difficulties, a mixture model~\cite{ME07}, a stochastic block model~\cite{PKS83} and their various extensions and combinations~\cite{BM11,TYSYR11,EDSE08,HXJ11,Peixoto14,AFL11} have been recently introduced to enable an \lq\lq exploratory\rq\rq~analysis of networks, allowing us to extract unspecified structural patterns even if some edges in the networks are missing~\cite{ACM08,RM09}. By fitting the model to the observed network structure, vertices with the same connection profiles are categorized into a predefined number of groups. The philosophy of these approaches is quite similar to that of the \lq\lq role model\rq\rq~in sociology~\cite{FH71}---individuals having locally or globally analogous relationships with others play the same \lq\lq role\rq\rq~or take up the same \lq\lq position\rq\rq~\cite{JD07}. It is clear to see that the possible topologies of the groups include community structure and multipartite structure, but they can be much, much wider.

One common assumption shared by these models is that the target networks contain positive links only. However, we frequently encounter the signed networks, which have both positive and negative
edges, in biology~\cite{CSA11,MGKQS09}, computer science~\cite{GGC12}, and last but definitely not
least, social science~\cite{BWJ07,SPA09,VJ09,MRS10}. The negative connections usually represent
hostility, conflict, opposition, disagreement, and distrust between individuals or organizations,
as well as the anticorrelation among objectives, whose coupled relation with positive links has been empirically shown to play a crucial role in the function and evolution
of the whole network~\cite{MGKQS09,MRS10}.

Several works have been conducted to detect community structure in these kinds of networks. Yang
\textit{et al.}~\cite{BWJ07} proposed an agent-based method that performs a random walk from one
specific vertex for a few steps to mine the communities in positive and signed networks. G\'{o}mez
\textit{et al.}~\cite{SPA09} presented a generalization of the widely-used
modularity~\cite{Newman06,MM04} to allow for negative links. Traag and Bruggeman~\cite{VJ09}
extended the Potts model to incorporate negative edges, resulting in a method similar to the
clustering of signed graphs. These approaches focus on the problem of community detection and thus they inevitably suffer a devastating failure if the signed networks
comprise other structural patterns, for example the disassortative structure, as shown in Sec.
\ref{sec4:sub1}. To make matters worse, they simply give a \lq\lq hard\rq\rq~partition
of signed networks in which a specific vertex could belong to one and only one cluster.
Similar to the positive networks, we have good reason to believe that the signed networks also simultaneously include all kinds of mesoscopic structures that might overlap with each other.

In this paper, we aim to capture and extract the intrinsic mesoscopic structure of networks with both positive and negative links. This goal is achieved by dividing the vertices into groups such that the vertices within each group have similar positive and negative connection patterns to other groups. We propose a generalized stochastic block model, referred to as signed stochastic block model (SSBM), in which the group memberships of each vertex are represented by unobserved or hidden quantities, and the relationship among groups is explicitly characterized by two block matrices, one for the positive links and the other for the negative links. By using
the expectation-maximization algorithm, we fit the model to the observed network structure and reveal the structural patterns without prior knowledge of what specific structures existing in the network. As a result, not only can various unspecific structures be successfully found, but also their types can be immediately elucidated by the block matrices. In addition, the model parameters tell us the fuzzy group memberships and the centrality of each vertex, which enable us to discover the networks' overlapping structures and to identify two kind of important vertices, i.e., group core and bridge. Experiments on a number of synthetic and real world networks validate the effectiveness and the advantage of our model.

The rest of this paper is organized as follows. We begin with the depictions of the mesoscopic structures, especially the definitions of the community structure and disassortative structure, in signed networks in Sec.~\ref{sec2}. Then we introduce an extension of the stochastic block model in Sec.~\ref{sec3}, and show how to employ it to perform an exploratory analysis of a given network with both positive and negative links. Experimental results on a series of synthetic networks with various designed structures and three social networks are given in Sec.~\ref{sec4}, followed by the conclusions in Sec.~\ref{sec5}.

\section{Mesoscopic structures in signed networks}
\label{sec2}

It is well known that the mesoscopic structural patterns in positive networks can be roughly classified into the following two different types: \lq\lq Assortative structure\rq\rq, usually called \lq\lq community structure\rq\rq~in most cases, refers to groups of vertices within which connections are relatively dense and between which they are sparser~\cite{MM02,Newman06,ME07}. In contrast, \lq\lq disassortative structure\rq\rq, also named \lq\lq bipartite structure\rq\rq~or more generally \lq\lq multipartite structure\rq\rq, means that network vertices have most of their connections outside their group~\cite{PFCB03,Newman06,ME07}.

For a signed network, its mesoscopic structure is quite different from and much more complicated
than that in a positive network since both the density and the sign of the links should be taken
into account at the same time. The intuitive descriptions of the assortative structure and
disassortative structure given in Ref.~\cite{Newman06,ME07} are no longer suitable. A natural question arises: How can we characterize the mesoscopic structures in a network that has both positive and negative edges? Guidance can be provided by the social balance theory~\cite{Heider46}, which states that the attitudes
of two individuals toward a third person should match if they are positively related. In this
situation, the triad is said to be socially balanced. A network is called balanced provided that
all its triads are balanced. This concept can be further generalized to
$k$-balance~\cite{Davis67,DF68} when the network can be divided into $k$ clusters, each having
only positive links within itself and negative links with others.

Following the principle, we can reasonably describe the community structure in a signed network
as a set of groups of vertices within which positive links are comparatively dense and negative
links are sparser, and on the contrary between which positive links are much looser and negative
links are thicker~\cite{BWJ07,VJ09,SPA09}. Obviously, it is an extension of the standard community
structure in networks with positive edges. In contrast, the disassortative structure
can be defined as a collection of vertices that have most of their negative links within the
group to which they belong while have majority of their positive connections outside their group.

\section{Methods}
\label{sec3}

\subsection{The SSBM Model}
\label{sec3:sub1}

Given a directed network $G = (V, E)$, we can represent it by an adjacency matrix $A$. The entries
of the matrix are defined as: $A_{ij} = 1$ if a positive link is present from vertex $i$ to vertex
$j$, $A_{ij} = -1$ if a negative link is present from vertex $i$ to vertex $j$, and $A_{ij} = 0$
otherwise. For weighted networks, $A_{ij}$ can be generalized to represent the weight of the link.
We further separate the positive component from the negative one by setting $A^+_{ij} = A_{ij}$ if
$A_{ij} > 0$ and $0$ otherwise, and $A^-_{ij} = -A_{ij}$ if $A_{ij} < 0$ and $0$ otherwise, so $A
= A^+ - A^-$.

Suppose that the vertices fall into $c$ groups whose memberships are \lq\lq hidden\rq\rq~or \lq\lq
missing\rq\rq~for the moment and will be inferred from the observed network structure. The
number of groups $c$ can also be inferred from the data, which will be discussed in Sec.
\ref{sec3:sub3}, but we take it as a given here. The standard solution for such an inference
problem is to give a generative model for the observed network structure and then to determine the
parameters of the model by finding its best fit~\cite{ME07,BM11,TYSYR11,EDSE08,HXJ11}.

\begin{figure}
\centering
\includegraphics[width = 0.45\textwidth]{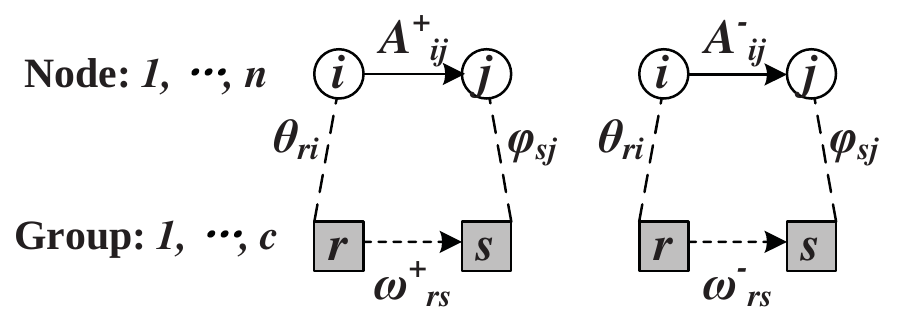}
\caption{Stochastic block model for signed networks. Unfilled circles represent observed network
structure and filled ones correspond to hidden memberships. The
solid line between vertex $i$ and $j$ indicates the existence of one positive or negative edge
connecting them. The dashed line indicates that the relation between the corresponding quantities
is unobserved and requires being learned from the observed network data.}
\label{fig1}
\end{figure}

The model we use is a kind of stochastic block model that parameterizes the probability of each
possible configuration of group assignments and edges as follows (see Fig.~\ref{fig1} for a
schematic illustration). Given an edge $e_{ij}$, we choose a pair of group $r$ and $s$ for its
tail and head with probability $\omega^+_{rs}$ if $e_{ij}$ is positive, or with probability
$\omega^-_{rs}$ if $e_{ij}$ is negative. The two scalars $\omega^+_{rs}$ and $\omega^-_{rs}$
giving the probability that a randomly selected positive and negative edge from group $r$ to $s$
respectively, explicitly characterize various types of connecting patterns among groups, as we
will see later. Then, we draw the tail vertex $i$ from group $r$ with probability $\theta_{ri}$
and the head vertex $j$ from group $s$ with probability $\phi_{sj}$. Intuitively, the parameter
$\theta_{ri}$ captures the centrality
of vertex $i$ in the group $r$ from the perspective of outgoing
edges while $\phi_{sj}$ describes the centrality of vertex $j$
in the group $s$ from the perspective of incoming edges. The parameters $\omega^+_{rs}$,
$\omega^-_{rs}$, $\theta_{ri}$ and $\phi_{sj}$ satisfy the normalization condition
\begin{eqnarray}
\sum_{r=1}^c \sum_{s=1}^c \omega^+_{rs} = 1, \quad \sum_{r=1}^c \sum_{s=1}^c \omega^-_{rs} = 1,
\nonumber\\
\sum_{i=1}^n \theta_{ri} = 1, \quad \sum_{j=1}^n \phi_{sj} = 1. \nonumber
\end{eqnarray}
Let $\lar{g}_{ij}$ and $\rar{g}_{ij}$ to be respectively the group membership of the tail and head
of the edge $e_{ij}$. So far, we have introduced all the quantities in our model: observed
quantities $\{A_{ij}\}$, hidden quantities $\{\rar{g}_{ij}, \lar{g}_{ij}\}$ and model parameters
$\{\omega^+_{rs}, \omega^-_{rs}, \theta_{ri}, \phi_{sj}\}$. To simplify the notations, we shall
henceforth denote by $\omega^+$ the entire set $\{\omega^+_{rs}\}$ and similarly $\omega^-$,
$\theta$, $\phi$, $\lar{g}$ and $\rar{g}$ for $\{\omega^-_{rs}\}$, $\{\theta_{ri}\}$,
$\{\phi_{sj}\}$, $\{\lar{g}_{ij}\}$ and $\{\rar{g}_{ij}\}$. The probability that we observe a
positive edge $e^+_{ij}$ can be written
as
\begin{equation}
\text{Pr}(e^+_{ij} | \omega^+, \theta, \phi) = \sum_{rs} \omega^+_{rs} \theta_{ri} \phi_{sj},
\label{eq1}
\end{equation}
and the probability of observing a negative edge $e^-_{ij}$ is
\begin{equation}
\text{Pr}(e^-_{ij} | \omega^-, \theta, \phi) = \sum_{rs} \omega^-_{rs} \theta_{ri} \phi_{sj}.
\label{eq2}
\end{equation}
The marginal likelihood of the signed network, therefore, can be represented by
\begin{eqnarray}
\text{Pr}(A|\omega^+, \omega^-, \theta, \phi)~~~~~~~~~~~~~~~~~~~~~~~~~~~~~~~~~~~~~~~\nonumber\\
~~~= \prod_{ij} \bigg(\sum_{rs} \omega^+_{rs}\theta_{ri}\phi_{sj}\bigg)^{A^+_{ij}} \bigg(\sum_{rs}
\omega^-_{rs}\theta_{ri}\phi_{sj}\bigg)^{A^-_{ij}}. \label{eq3}
\end{eqnarray}
Note that the self-loop links are allowed and the weight $A^+_{ij}$ and $A^-_{ij}$ are
respectively viewed as the number of positive and negative multiple links from vertex $i$ to
vertex $j$ as done in many existing models~\cite{TYSYR11,EDSE08,HXJ11}.

To infer the missing group memberships $\lar{g}$ and $\rar{g}$, we need to maximize the likelihood
in Eq.~(\ref{eq3}) with respect to the model parameters $\omega^+$, $\omega^-$, $\theta$ and
$\phi$. For convenience, one usually works not directly with the likelihood itself but with its
logarithm
\begin{eqnarray}
\mathcal{L}&=& \text{ln}\text{Pr}(A|\omega^+, \omega^-, \theta,
\phi)\nonumber \\
&=&
\sum_{ij}{A^+_{ij}\text{ln}\left(\sum_{rs}\omega^+_{r,s}\theta_{ri}\phi_{sj}\right)} \nonumber\\
& &+ \sum_{ij}{A^-_{ij}\text{ln}\left(\sum_{rs}\omega^-_{r,s}\theta_{ri}\phi_{sj}\right)}.
\label{eq4}
\end{eqnarray}
The maximum of the likelihood and its logarithm occur in the same place because the
logarithm is a monotonically increasing function.

Considering that the group memberships $\lar{g}$ and $\rar{g}$ are unknown, it is intractable to
optimize the log-likelihood $\mathcal{L}$ directly again. We can, however, give a good guess of
the hidden variables $\lar{g}$ and $\rar{g}$ according to the network structure and the model
parameters, and seek the maximization of the following expected log-likelihood
\begin{widetext}
\begin{eqnarray}
\overline{\mathcal{L}}&=&\sum_{\lar{g},\rar{g}}\text{Pr}(\lar{g},\rar{g}|A^+,\omega^+,
\theta,\phi) \text{ln}\text{Pr}(A^+|\lar{g},\rar{g}, \omega^+, \theta, \phi)
+ \sum_{\lar{g},\rar{g}}\text{Pr}(\lar{g},\rar{g}|A^-,\omega^-, \theta,\phi)
\text{ln}\text{Pr}(A^-|\lar{g},\rar{g}, \omega^-, \theta, \phi)\nonumber\\
&=&\sum_{ijrs}\text{Pr}(r,s|e^+_{ij},\omega^+,\theta, \phi)
\bigg[A^+_{ij}\big(\text{ln}\omega^+_{rs}+\text{ln}\theta_{ri}+\text{ln}\phi_{sj}\big)\bigg]
+ \sum_{ijrs}\text{Pr}(r,s|e^-_{ij},\omega^-,\theta, \phi)
\bigg[A^-_{ij}\big(\text{ln}\omega^-_{rs}+\text{ln}\theta_{ri}+\text{ln}\phi_{sj}\big)\bigg]
\nonumber \\
&=&\sum_{ijrs}{q^+_{ijrs}A^+_{ij}\left(\text{ln}\omega^+_{rs}+\text{ln}\theta_{ri}+\text{ln}\phi_{sj}\right)}
+
\sum_{ijrs}{q^-_{ijrs}A^-_{ij}\left(\text{ln}\omega^-_{rs}+\text{ln}\theta_{ri}+\text{ln}\phi_{sj}\right)},
\label{eq5}
\end{eqnarray}
\end{widetext}
where $q^+_{ijrs} = \text{Pr}(\lar{g}_{ij} = r, \rar{g}_{ij} = s|e^+_{ij}, \omega^+, \theta,
\phi)$ is the probability that one find a positive edge $e^+_{ij}$ with its tail vertex $i$ from
group $r$ and its head vertex $j$ from group $s$ given the network and the model parameters.
Analogous interpretation can be made for $q^-_{ijrs} = \text{Pr}(\lar{g}_{ij} = r, \rar{g}_{ij} =
s|e^-_{ij}, \omega^-, \theta, \phi)$ too.

With the expected log-likelihood, we can get the best estimate of the value of $\mathcal{L}$
together with the position of its maximum gives the most likely values of the model parameters.
Finding the maximum still presents a problem, however, since the calculation of $q^+_{ijrs}$ and
$q^-_{ijrs}$ requires the values of $\omega^+$, $\omega^-$, $\theta$ and $\phi$, and vice versa.
One possible solution is to adopt an iterative self-consistent approach that evaluates both
simultaneously. Like many previous works~\cite{ME07,TYSYR11,EDSE08,HXJ11}, we utilize the
expectation-maximization (EM) algorithm, which first computes the posterior probabilities of
hidden variables using estimated model parameters and observed data (the E-step), and then
re-estimates the model parameters (the M-step).

In the E-step, we calculate the expected probabilities $q^+_{ijrs}$ and $q^-_{ijrs}$ given the
observed network $A$ and parameters $\omega^+$, $\omega^-$, $\theta$ and $\phi$
\begin{equation}
\left\{
\begin{aligned}
q^+_{ijrs} &=& \frac{\text{Pr}(\lar{g}_{ij} = r, \rar{g}_{ij} = s, e^+_{ij} | \omega^+, \theta,
\phi)}{\text{Pr}(e^+_{ij}|\omega^+, \theta, \phi)} \\
&=& \frac{\omega^+_{rs}\theta_{ri}\phi_{sj}}{\sum_{rs}
\omega^+_{rs}\theta_{ri}\phi_{sj}},~~~~~~~~~~~~~~~~~~~~~~~~ \\
q^-_{ijrs} &=& \frac{\text{Pr}(\lar{g}_{ij} = r, \rar{g}_{ij} = s, e^-_{ij} | \omega^-, \theta,
\phi)}{\text{Pr}(e^-_{ij}|\omega^-, \theta, \phi)} \\
&=& \frac{\omega^-_{rs}\theta_{ri}\phi_{sj}}{\sum_{rs}
\omega^-_{rs}\theta_{ri}\phi_{sj}}.~~~~~~~~~~~~~~~~~~~~~~~~
\end{aligned} \right.
\label{eq6}
\end{equation}
In the M-step, we use the values of $q^+_{ijrs}$ and $q^-_{ijrs}$ estimated in the E-step, to
evaluate the expected log-likelihood and to find the values of the parameters that maximize it.
Introducing the Lagrange multipliers $\rho^+$, $\rho^-$, $\gamma_r$ and $\lambda_s$ to incorporate
the normalization conditions, the expected log-likelihood expression to be maximized becomes
\begin{eqnarray}
\tilde{\mathcal{L}}&=&\overline{\mathcal{L}}+\rho^+\bigg(1-\sum_{rs}{\omega^+_{rs}}\bigg)
+\rho^-\bigg(1-\sum_{rs}{\omega^-_{rs}}\bigg)+\nonumber\\
&&
\sum_{r}{\gamma_{r}\bigg(1-\sum_{i}{\theta_{ri}}\bigg)}
+\sum_{s}{\lambda_{s}\bigg(1-\sum_{j}{\phi_{sj}}\bigg)}. \label{eq7}
\end{eqnarray}
By letting the derivative of $\tilde{\mathcal{L}}$ to be 0, the maximum
of the expected log-likelihood appears at the places where
\begin{equation}
\left\{
\begin{aligned}
\omega^+_{rs}&=\frac{ \sum_{ij}{A^+_{ij}q^+_{ijrs}} } { \sum_{ijrs}{A^+_{ij}q^+_{ijrs}}}, \\
\omega^-_{rs}&=\frac{ \sum_{ij}{A^-_{ij}q^-_{ijrs}} } { \sum_{ijrs}{A^-_{ij}q^-_{ijrs}}}, \\
\theta_{ri} &=
\frac{\sum_{js}{A^+_{ij}q^+_{ijrs}}+\sum_{js}{A^-_{ij}q^-_{ijrs}}}{\sum_{ijs}{A^+_{ij}q^+_{ijrs}}+\sum_{ijs}{A^-_{ij}q^-_{ijrs}}},
\\
\phi_{sj} &=
\frac{\sum_{ir}{A^+_{ij}q^+_{ijrs}}+\sum_{ir}{A^-_{ij}q^-_{ijrs}}}{\sum_{ijr}{A^+_{ij}q^+_{ijrs}}+\sum_{ijr}{A^-_{ij}q^-_{ijrs}}}.
\end{aligned} \right.
\label{eq8}
\end{equation}
Eq.~(\ref{eq6}) and (\ref{eq8}) constitute our EM algorithm for exploratory analysis of signed
networks. When the algorithm converges, we obtain a set of values for hidden quantities
$q^+_{ijrs}$, $q^-_{ijrs}$ and model parameters $\omega^+$, $\omega^-$, $\theta$ and $\phi$.

It is worthwhile to note that the EM algorithm are known to converge to local maxima of the
likelihood but not always to global maxima. With different starting values, the algorithm may give
rise to different solutions. To obtain a satisfactory solution, we perform several runs with
different initial conditions and return the solution giving the highest log-likelihood over all
the runs.

Now we consider the computational complexity of the EM algorithm. For each iteration, the cost
consists of two parts. The first part is from the calculation of $q^+_{ijrs}$ and $q^-_{ijrs}$ using Eq.~(\ref{eq6}), whose time complexity is $O(m\times c^2)$. Here $m$
is the edges in the network and $c$ is the number of groups. The
second part is from the estimation of the model parameters using
Eq.~(\ref{eq8}), whose time complexity is also $O(m\times c^2)$. We
use $T$ to denote the number of iterations before the iteration
process converges. Then, the total cost of the EM algorithm for our model is $O(T\times m
\times c^2)$. It is difficult to give a theoretical estimation to
the number $T$ of iterations. Generally speaking, $T$ is determined
by the network structure and the initial condition.

\subsection{Soft partition and overlapping structures}
\label{sec3:sub2}

The parameters, obtained by fitting the model to the observed network structure with the E-M
algorithm, provide us useful information for the mesoscopic structure in a given network.
Specifically, the matrices $\omega^+$ and $\omega^-$, an analogy with the image graph in the role
model~\cite{JS06}, characterize the connecting patterns among different groups, which determine
the type of structural patterns. Furthermore, $\theta$ and $\phi$ indicate the centrality of a
vertex in its groups from the perspective of outgoing edges and incoming edges, respectively.
Consequently, the probability of vertex $i$ drawn from group $r$ when it is the tail of edges can
be defined as
\begin{equation}
\alpha_{ir} = \frac{\sum_s (\omega^+_{rs} + \omega^-_{rs})\theta_{ri}}{\sum_{rs} (\omega^+_{rs} +
\omega^-_{rs})\theta_{ri}},
\label{eq9}
\end{equation}
and vertex $i$ can be simply assigned to the group $r^*$ to which it most likely belongs, i.e.,
$r^* = \arg\max_r \{\alpha_{ir}, r = 1, 2, \dots, c\}$. The result gives a hard partition of the
signed network.

In fact, the set of scalars $\{\alpha_{ir}\}_{r=1}^c$ supply us with the probabilities that vertex
$i$ belongs to different groups, which can be referred to as the soft or fuzzy memberships.
Assigning vertices to more than one group have attracted by far the most interest, particularly in
overlapping community detection~\cite{GIIT05,DJP05,TAF08,ISMB11}. The vertices belonging to
several groups, are found to take a special role in networks, for example, signal transduction in
biological networks. Furthermore, some vertices, considered as \lq\lq instable\rq\rq~\cite{DJP05},
locate on the border between two groups and thus are difficult to classify into any group. It is
of great importance to reveal the global organization of a signed network in terms of overlapping
mesoscopic structures and to find the instable vertices. We employ here the
bridgeness~\cite{TAF08} and group entropy~\cite{JL12} to capture the vertices' instabilities and
to extract the overlapping mesoscopic structure. These two measures of vertex $i$ are computed as
\begin{equation}
b_i = 1 - \sqrt{\frac{c}{c-1}\sum_{r=1}^c \bigg(\alpha_{ir} - \frac{1}{c}\bigg)^2},
\label{eq10}
\end{equation}
\begin{equation}
\xi_i = - \sum_{r=1}^c \alpha_{ir} \log_{c} \alpha_{ir}.
\label{eq11}
\end{equation}
Note that vertex $i$ has a large bridgeness $b_i$ and entropy $\xi_i$ when it most likely
participates in more than one group simultaneously and vice versa. From the perspective of
incoming edges, we can represent the probability of vertex $j$ belonging to group $s$ by
\begin{equation}
\beta_{js} = \frac{\sum_r (\omega^+_{rs} + \omega^-_{rs})\phi_{sj}}{\sum_{rs} (\omega^+_{rs} +
\omega^-_{rs})\phi_{sj}}.
\label{eq12}
\end{equation}
These statements for $\alpha_{ir}$ also apply to $\beta_{js}$. So we don't need to repeat again.

The model described above focus on directed networks. Actually, the model could be easily
generalized to undirected networks by letting the parameter $\theta$ be identical to $\phi$. The
derivation follows the case of directed networks and the results are the same to Eq.~(\ref{eq6})
and (\ref{eq8}).

\subsection{Model selection}
\label{sec3:sub3}

So far, our model assumes that the number of groups $c$ is known as a prior. This information,
however, is unavailable for many cases. It is necessary to provide a criterion to determine an
appropriate group number for a given network. Several methods have been proposed to deal with this
model selection issue. We adopt the minimum description length (MDL) principle, which is also
utilized in the previous generative models for network structure exploration~\cite{HXJ11}.

According to MDL principle, the required length to describe the network data comprises two
components. The first one describes the coding length of the network, which is $-L$ for directed
network and $-L/2$ for undirected network. The other gives the length for coding model parameters
that is $-\sum_{rs} \text{ln}\omega^+_{rs} - \sum_{rs} \text{ln}\omega^-_{rs} -\sum_{ri}\text{ln}\theta_{ri} - \sum_{sj}\text{ln}\phi_{sj}$ for directed network and $-\sum_{rs}
\text{ln}\omega^+_{rs} - \sum_{rs} \text{ln}\omega^-_{rs} -\sum_{ri}\text{ln}\theta_{ri}$ for
undirected network. The optimal $c$ is the one which minimizes the total description length.

\section{Experimental results}
\label{sec4}

In this section, we extensively test our SSBM model on a series of synthetic signed networks with
various known structure, including community structure and disassortative structure. After
that, the method is also applied to three real-life social networks.

\subsection{Synthetic networks}
\label{sec4:sub1}

The ad hoc networks, designed by Girvan and Newman~\cite{MM02}, have been broadly used to validate
and compare community detection algorithms~\cite{LJAA05,MM04,DJP05,JL12}. By contrast, there
exists no such benchmark for community detection in networks with both positive and negative
links. We generate the signed ad hoc networks with controlled community structure by the method
developed in Refs.~\cite{BWJ07,KF08}. The networks have 128 vertices, which are divided into four
groups with 32 vertices each. Edges are placed randomly such that they are positive within groups
and negative between groups, and the average degree of a vertex to be 16. The community structure
is controlled by three parameters, $p_\mathrm{in}$ indicating the probability of each vertex connecting
to other vertices in the same group, $p_+$ the probability of positive links appearing between
groups, and $p_-$ the probability of negative links arising within groups. Thus, the parameter
$p_\mathrm{in}$ regulates the cohesiveness of the communities and the remaining parameters $p_+$ and
$p_-$ add noise to the community structure when $p_{in}$ is fixed.

\begin{figure}[!t]
\centering
\includegraphics[width = 0.48\textwidth]{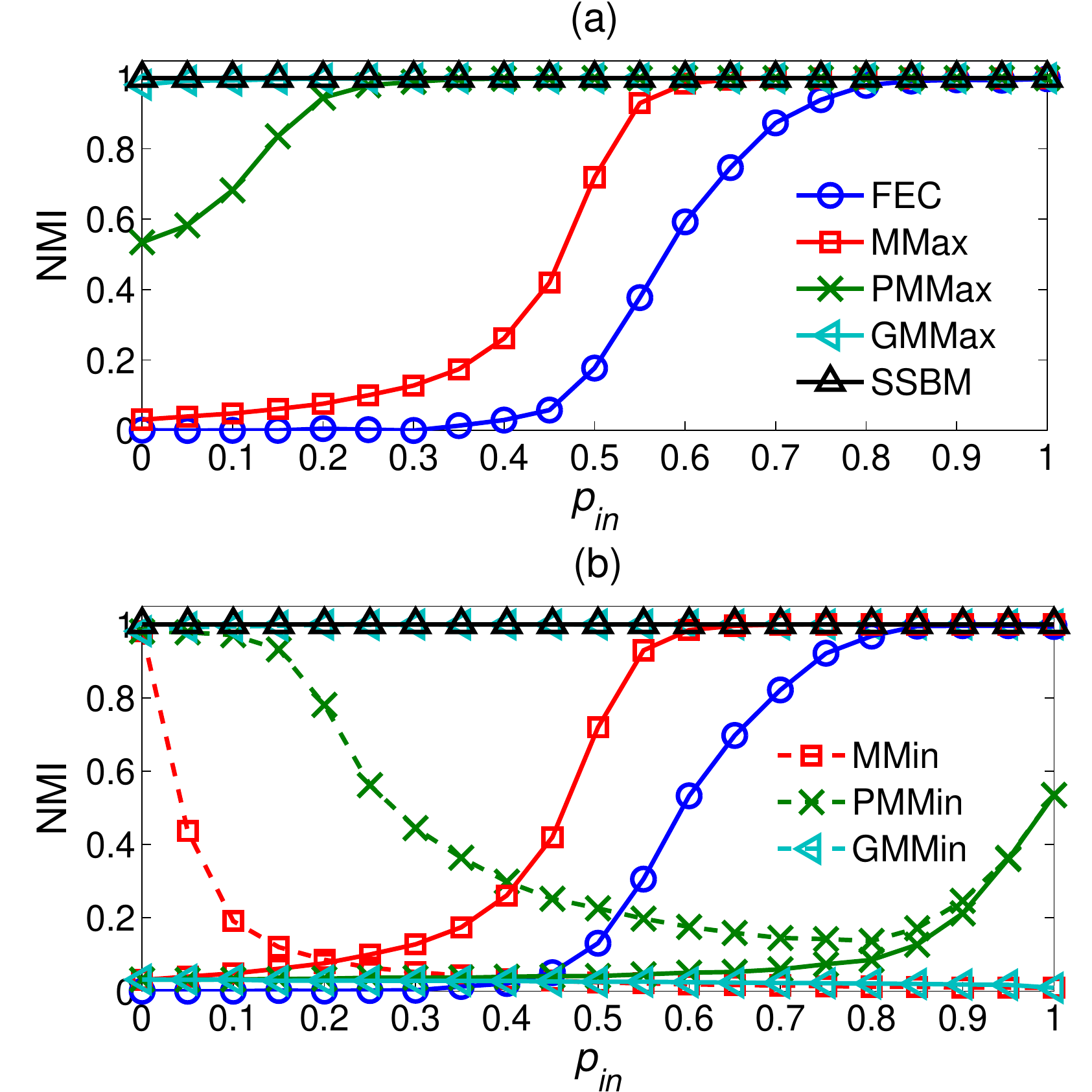}
\caption{(Color online) NMI of our method and other approaches on balanced ad-hoc
networks with controlled community structure (a) and disassortative structure (b). Each point is
an average over 50 realizations of the networks.}
\label{fig2}
\end{figure}

For the synthetic networks, we simply consider their hard partition as defined in
Sec.~\ref{sec3:sub2}. The results are evaluated by the normalized mutual information
(NMI)~\cite{SJ02}, which can be formulated as
\begin{equation}
\text{NMI}(C_1, C_2) = \frac{\sum\limits_{i=1}^c\sum\limits_{j=1}^c n_{ij}\text{ln}\frac{n_{ij}n}{n^{(1)}_i
n^{(2)}_j}}{\sqrt{(\sum\limits_{i=1}^c n^{(1)}_i \text{ln}\frac{n^{(1)}_i}{n})(\sum\limits_{i=1}^c n^{(2)}_i
\text{ln}\frac{n^{(2)}_i}{n})}} \nonumber
\end{equation}
where $C_1$ and $C_2$ are the true group assignment and the assignment found by the algorithms,
respectively, $n$ is the number of vertices, $n_{ij}$ is the number of vertices in the known group
$i$ that are assigned to the inferred group $j$, $n^{(1)}_i$ is the number of vertices in the true
group $i$, $n^{(2)}_j$ is the number of vertices in the inferred group $j$. The larger the NMI
value, the better the partition obtained by the algorithms.

\begin{figure*}[!t]
\centering
\includegraphics[width = 0.9\textwidth]{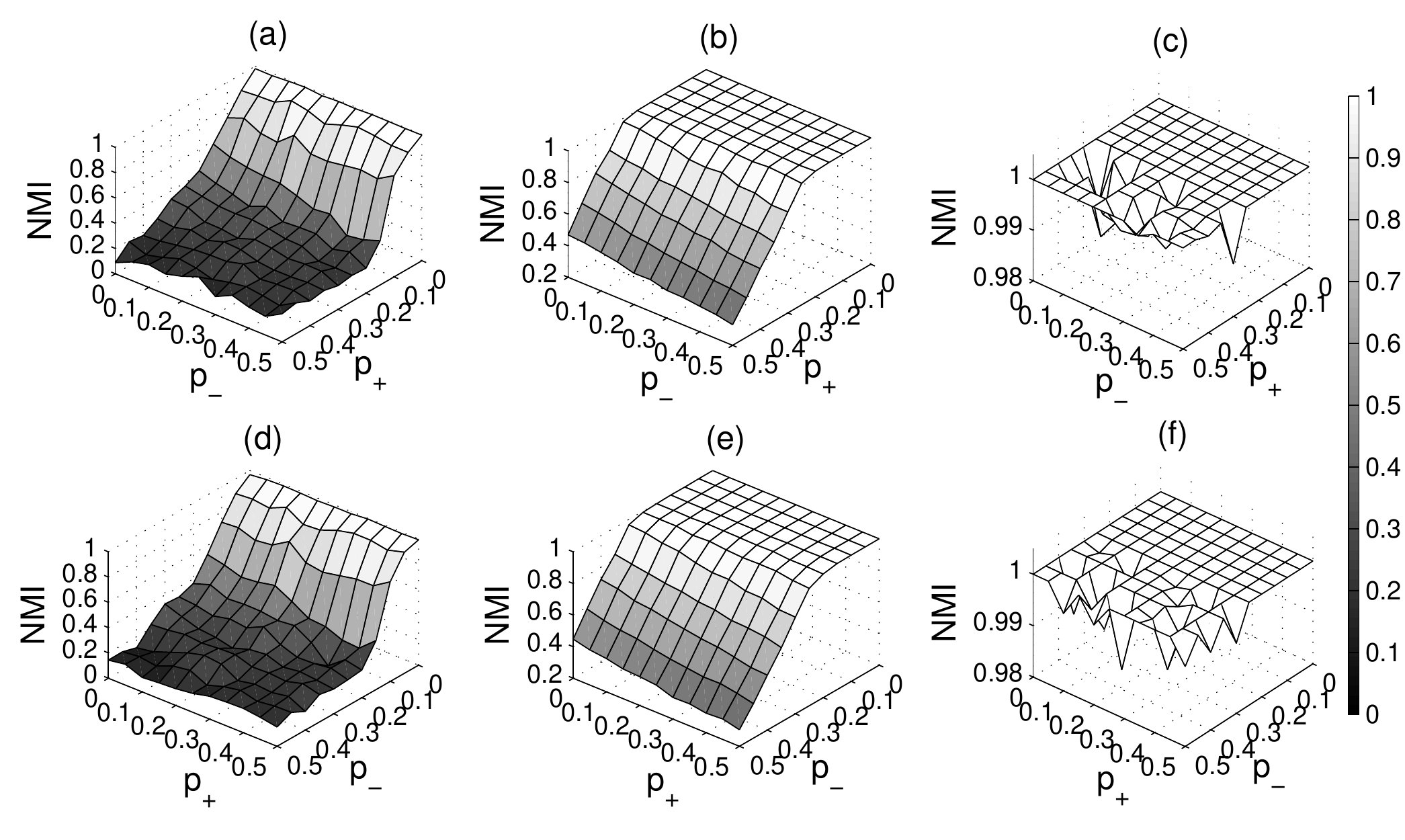}
\caption{NMI on unbalanced ad-hoc networks with controlled community
structure (first row) for (a) FEC, (b) GMMax and (c) SSBM,  and with controlled disassortative structure (second row) for (d) FEC, (e) GMMin and (f) SSBM. Each point is an average over 50 realizations of the networks.}
\label{fig3}
\end{figure*}

We conduct two different experiments. First, we set the two parameters $p_+$ and $p_-$ to be zero
and gradually change $p_\mathrm{in}$ from 1 to 0. In this situation, all the generated synthetic networks
are 4-balanced. Fig.~\ref{fig2}(a) reports the experimental results obtained by our method and two
state-of-the-art approaches, namely generalized modularity maximization through simulated annealing
(denoted by GMMax)~\cite{VJ09,SPA09} and the finding and extracting community (FEC) method~\cite{BWJ07}. In addition, we also
implement the simulated annealing algorithm to maximize the standard modularity by ignoring the
sign of the links (denoted by MMax) and removing the negative edges (denoted by PMMax),
respectively. Each point in the curves is an average over 50 realization of the synthetic random
networks. Bear in mind that the community structure becomes less cohesive as the parameter
$p_\mathrm{in}$ decreases from 1 to 0. We can see that both the SSBM model and the GMMax method perform fairly well and are almost able to perfectly recover the communities in the synthetic networks for
all cases. When $0 \leq p_\mathrm{in} \leq 0.1$, our model is even slightly superior to the GMMax
approach. The remaining three methods, however, can only achieve promising results when $p_\mathrm{in}$ is
sufficiently large. They all show a fast deterioration as $p_\mathrm{in}$ becomes smaller and
smaller. For example, the NMI of the FEC algorithm begins to drop once $p_{in}$ exceeds 0.8, and
then quickly reduces to less than 0.2 when $p_\mathrm{in} = 0.5$ and even to approximately 0 when
$p_\mathrm{in}$ is smaller than 0.3. Similar performances can be observed for the MMax and PMMax
approaches as well. These results are quite understandable since both the SSBM model and the GMMax
method consider the contribution made by the negative links in signed networks, which is either
neglected or removed in the remaining three approaches. This highlights the importance of the
negative edges for community detection in the signed networks. Moreover, the PMMax method always
outshines the MMax method, especially when $p_\mathrm{in}$ in the range $0 \leq p_\mathrm{in} \leq 0.5$, which
is in agreement with the results reported in Ref.~\cite{KF08}, indicating that the positive links
in signed networks have a significant impact on community detection.

Then, we fix the parameter $p_\mathrm{in} = 0.8$ and gradually change other two parameters $p_+$ and
$p_-$ from 0 to 0.5, respectively. Clearly, all the synthetic networks are not balanced in this
setting. The results obtained by our model and two updated algorithms are give in the upper row of
Fig~\ref{fig3}. As we can see, the SSBM model consistently, and sometimes significantly, outperforms
the other two approaches. More specifically, its NMF is always 1 expect for a few negligible
perturbations. By contrast, the FEC algorithm cannot offer a satisfactory partition of the signed
networks when $0 \leq p_+ \leq 0.3$ and $0 \leq p_- \leq 0.5$, whose NMI is less than 0.4 at all
times. When $0.3 \leq p_+ \leq 0.5$ and $0 \leq p_- \leq 0.5$, the GMMax approach exhibits a
competitive performance, but its NMI suddenly collapses and continuously decreases
once $p_+$ is larger than 0.3.

We turn now to the second experiment in which the synthetic networks have the controlled
disassortative structure. The signed networks are generated in the same way, expect that we
randomly place negative links within groups and positive links between groups. Similarly,  the
disassortative structure in these networks are controlled by three parameters again. $p_\mathrm{in}$
indicates the probability of each vertex connecting to other vertices in the same group, $p_+$ the
probability of positive links appearing within groups, and $p_-$ the probability of negative links
arising between groups.

We first study the balanced networks by setting $p_+$ and $p_-$ to be zero and changing $p_\mathrm{in}$
from 1 to 0 once again. As shown in Fig.~\ref{fig2}(b), the FEC algorithm, the MMax method and our
model achieve the performances that is very similar to those in the first experiment. That is, our
model always successfully find the clusters in the synthetic networks for all the cases, while
the FEC algorithm and the MMax method perform fairly well when $p_\mathrm{in}$ is large enough, but
quickly degrade as $p_\mathrm{in}$ approaches 0. The PMMax and the GMMax methods, however,
perform rather badly. The NMI of the PMMax method seems no greater than 0.5 even if $p_\mathrm{in} = 1$,
while the NMI of the GMMax approach nearly vanishes for all the cases. This is because the two methods, which seek standard and generalized modularity maximization, respectively, are suitable only for community detection. As a consequence, they deserve to suffer a serious failure in this experiment. Instead, one should minimize the modularity to uncover the multipartite structure in networks, as indicated in Ref.~\cite{Newman06}. Therefore, we apply the simulated annealing algorithm to minimize the generalized modularity (denoted by GMMin) and the standard modularity by ignoring the
sing of links (denoted by MMin) and excluding the negative connections (denoted by PMMin),
respectively. We see from Fig.~\ref{fig2}(b) that the GMMin method can obtain competitive
performance with our SSBM model expect for a slight inferior when $0 \leq p_\mathrm{in} \leq 0.1$. However, the MMin and the PMMin approaches perform unsatisfactorily due to the fact that they do not consider the contributions derived from the negative links.

We investigate next the disassortative structure in unbalanced synthetic networks by
fixing $p_\mathrm{in} = 0.8$ and changing $p_+$ and $p_-$ from 0 to 0.5 step by step. The lower row of
Fig.~\ref{fig3}~gives the results obtained by the FEC method, the GMMin approach and our SSBM
model, which are quite similar to those in the first experiment. In particular, although the SSBM does not
perform perfectly in some cases, its NMF is still rather high, say, more than 0.98. When $0
\leq p_- \leq 0.3$, the GMMin approach yields sufficiently good results, but its NMF reduces at a very fast speed along with $p_-$ toward 0.5. The FEC algorithm achieves the worst performance in all cases.

\begin{figure*}[!t]
\centering
\includegraphics[width = 0.9\textwidth]{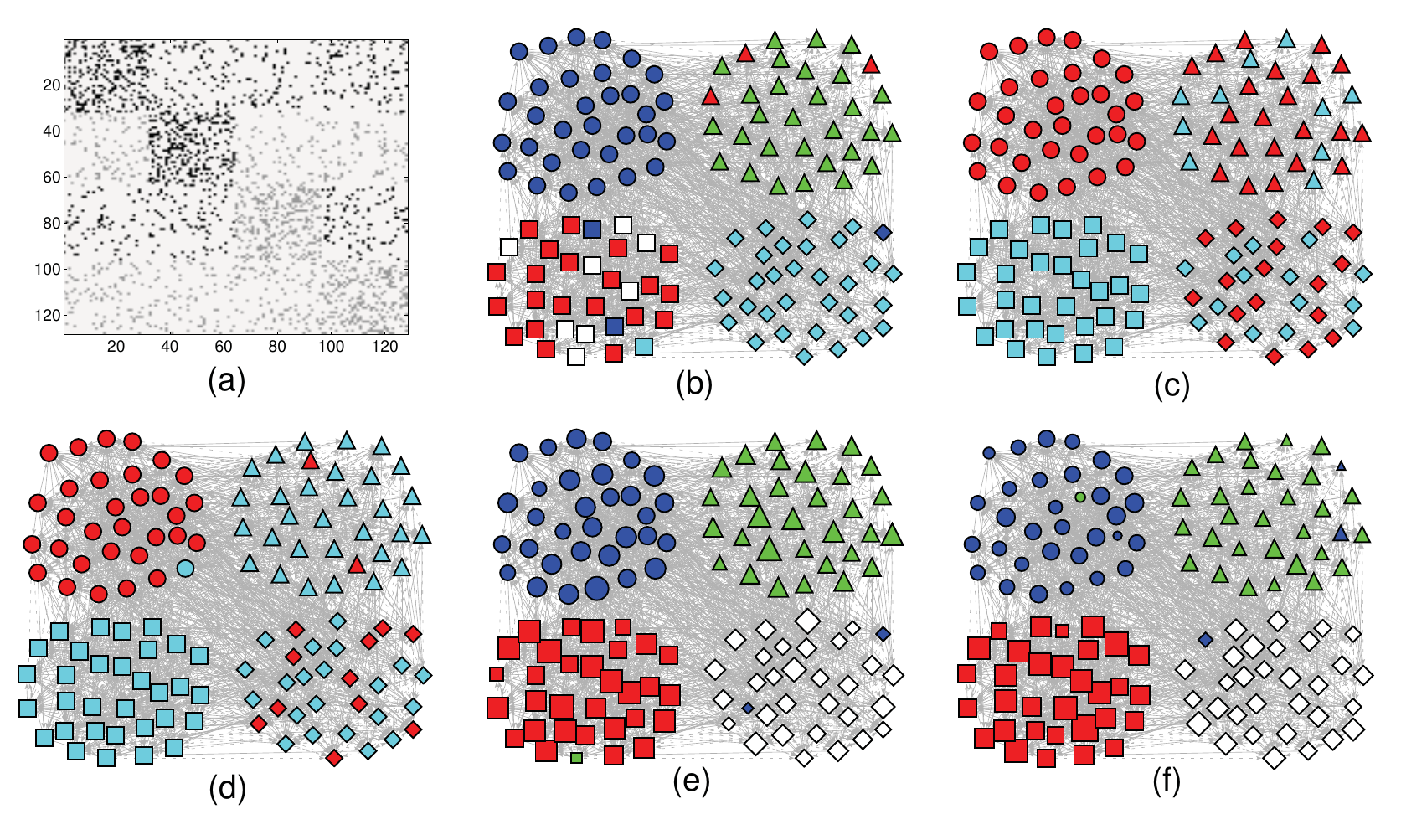}
\caption{(Color online) Detecting the mesoscopic structure of a synthetic network. (a) The adjacency matrix of the signed network where the black dots denote the positive links and
the gray dots represent the negative edges. The partitioning results for different methods (b) EFC, (c) GMMax, (d) GMMin and SSBM from the perspective of outgoing edges (e) and incoming edges (f), where the solid edges denote the positive links and the dashed edges represent negative links. The sizes of the vertices in (e) and (f) indicate their centrality degree in the corresponding groups according to the parameters $\theta$ and $\phi$, respectively.}
\label{fig4}
\end{figure*}

Finally, we focus on a synthetic network containing a multitude of mesoscopic structures, whose
adjacency matrix is given in Fig.~\ref{fig4}(a). Intuitively, according to the outgoing
edges in this network, the second group is the community structure and the third group belongs to
the disassortative structure. The first group with positive outgoing links only, can be viewed as
an example of the standard community structure in positive networks, while the last group, which includes only negative outgoing links, can be referred to as an extreme example of the disassortative structure in signed networks. Meanwhile, from the perspective of incoming edges, the four groups exhibit different types of structural patterns, which cannot be categorized simply as community structure or disassortative structure. We apply the FEC algorithm, the GMMax method, the GMMin method and our model to this signed network. Limited by their intrinsic assumptions, the FEC algorithm, the GMMax method and the GMMin method fail to uncover the structural patterns, as shown in Fig.~\ref{fig4}(b)-(d). In particular, the generalized modularity proposed in Refs.~\cite{VJ09,SPA09}, regardless of whether it is maximum or minimum, misleads us into receiving an improper partition of the network in which the four groups merge with each other. But by dividing vertices with the same connection profiles into groups, our model could accurately detect all types of mesoscopic structures, both from the perspective of outgoing links (Fig.~\ref{fig4}(e)) and from the perspective of incoming edges (Fig.~\ref{fig4}(f)). Furthermore, the obtained parameters $\theta$ and $\phi$ reveal the centrality of each vertex in its corresponding group from the two perspectives.

\begin{figure}[!t]
\centering
\includegraphics[width = 0.48\textwidth]{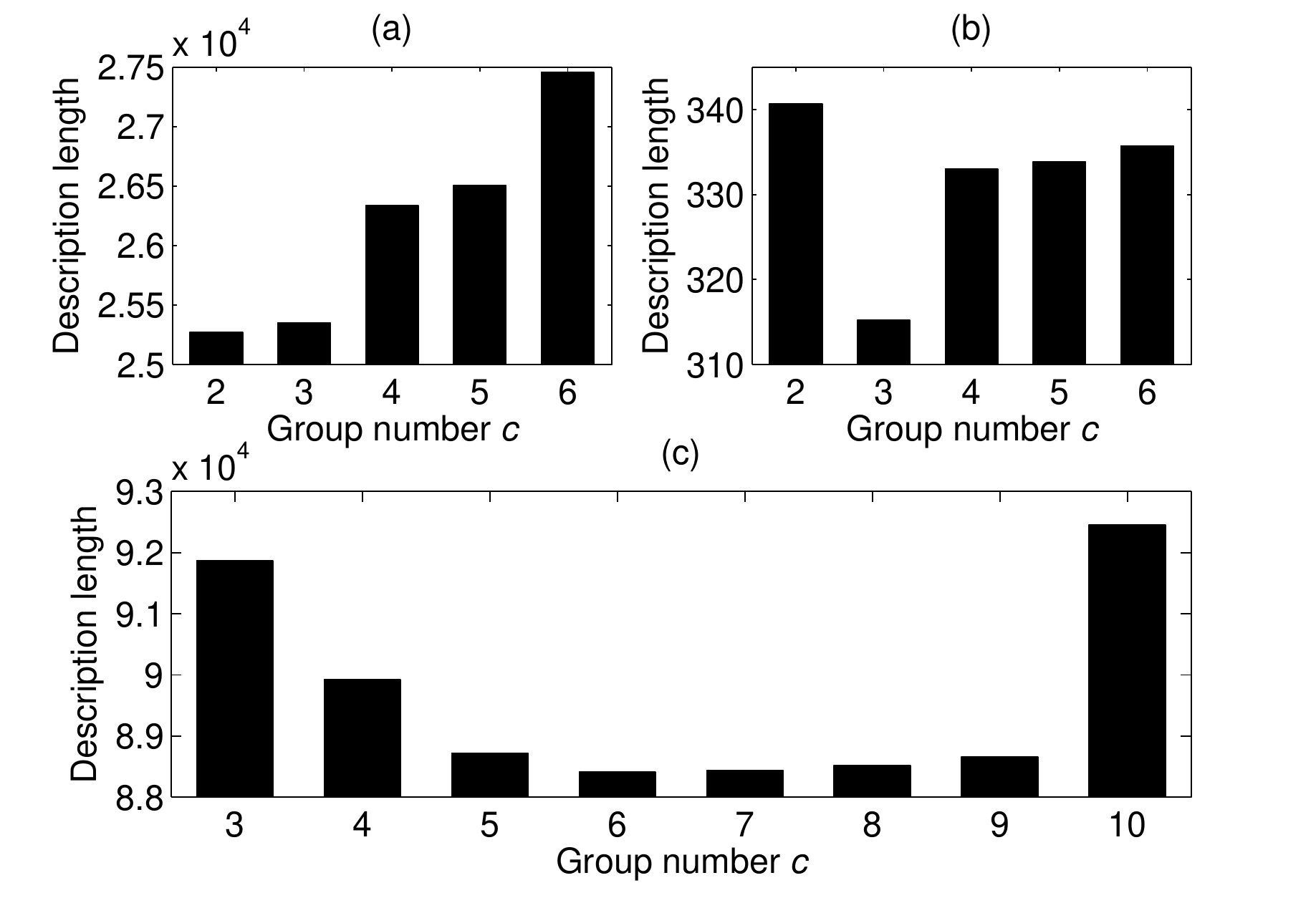}
\caption{Model selection results for (a) the Slovene Parliamentary network,
(b) the Gahuku-Gama Subtribes network and (c) the international conflict and alliance network.}
\label{fig8}
\end{figure}

\subsection{Real-life networks}
\label{sec4:sub2}

We further test our method by applying it to several real networks containing
both positive and negative links. The first network is a relation graph of 10 parties of the
Slovene Parliamentary in 1994~\cite{KM96}. The weights of links in the network were estimated by 72 questionnaires among 90 members of the Slovene National Parliament.
The questionnaires were designed to estimate the distance of the ten parties on a scale from -3 to
3, and the final weights were the averaged values multiplied by 100.

\begin{table*}[!t]
\centering
\caption{The soft group membership $\alpha$, bridgeness $b_i$~\cite{TAF08} and group entropy
$\xi_i$~\cite{JL12} of each vertex in the Slovene Parliamentary
network~\cite{KF08}. Larger bridgeness or entropy means that the corresponding node are more
\lq\lq instable\rq\rq.}
\vspace{2mm}
\normalsize
\begin{tabular}{ccccccccccc}
\hline
\hline
 Vertex & SKD&	ZLSD & SDSS & LDS & ZS-ESS & ZS & DS & SLS& SPS-SNS & SNS\\
\hline
 $\alpha_{i1}$ & 1.000 & 0 & 1.000 & 0	& 0 & 1.000 & 0 & 1.000 & 1.000 & 0.0186\\
 $\alpha_{i2}$ & 0 & 1.000 & 0 & 1.000 & 1.000 & 0 & 1.000 & 0 & 0 & 0.9814\\
 $b_i$ & 0 & 0 & 0 & 0 & 0 & 0 & 0 & 0 & 0 & 0.0372\\
 $\xi_i$& 0 & 0 & 0 & 0 & 0 & 0 & 0 & 0 & 0 & 0.1334 \\
\hline
\hline
\end{tabular}
\label{tab1}
\end{table*}

\begin{figure}[!t]
\centering
\includegraphics[width = 0.45\textwidth]{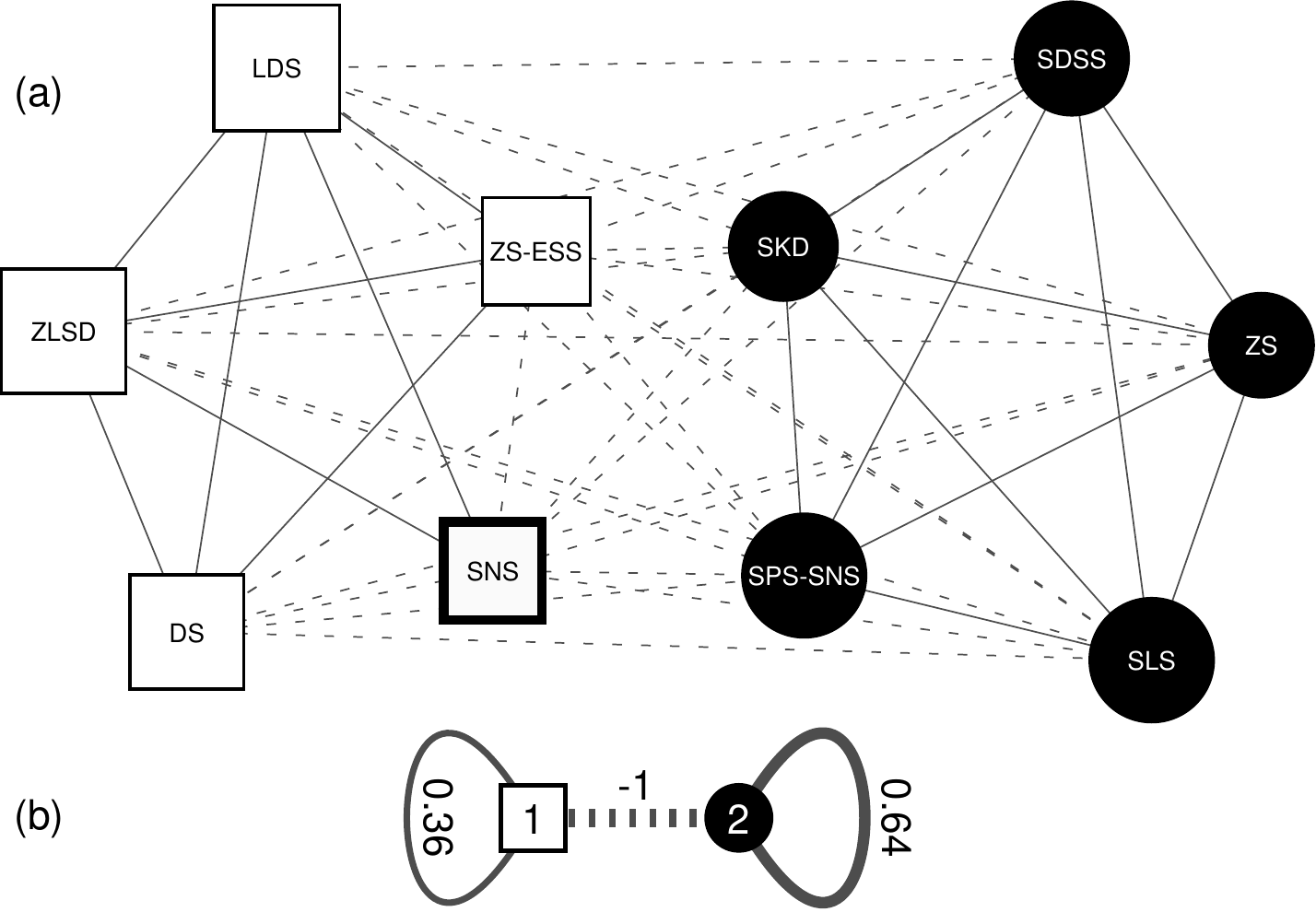}
\caption{Exploratory analysis of the Slovene Parliamentary network~\cite{KM96}. The solid edges denote the positive links and the dashed edges represent negative links. The true community
structure in this network is represented by two different shapes, circle and square. The shades of
nodes indicate the membership $\alpha$ obtained by fitting our model to this network. The sizes of
the vertices, proportional to $\theta$, indicates their centrality degree with respect to their
corresponding group.}
\label{fig5}
\end{figure}

We further test our method by applying it to several real networks containing
both positive and negative links. The first network is a relation graph of 10 parties of the
Slovene Parliamentary in 1994~\cite{KM96}. The weights of links in the network were estimated by 72 questionnaires among 90 members of the Slovene National Parliament.
The questionnaires were designed to estimate the distance of the ten parties on a scale from -3 to
3, and the final weights were the averaged values multiplied by 100.

\begin{table*}[!t] \caption{The soft group membership $\alpha$, bridgeness $b_i$~\cite{TAF08} and
group entropy $\xi_i$~\cite{JL12} of each vertex in the Gahuku-Gama Subtribes
network~\cite{Read54}. Larger bridgeness or entropy means that the corresponding node are more \lq\lq instable\rq\rq.}
\vspace{2mm}
\begin{tabular}{ccccccccccc}
\hline
\hline
Vertex & GAVEV & KOTUN & OVE & ALIKA & NAGAM & GAHUK & MASIL & UKUDZ & NOTOH & KOHIK\\
\hline
$\alpha_{i1}$ & 1.000 &	1.000 &	0 & 0 & 0 & 0 & 0 & 0 & 0 & 0\\
$\alpha_{i2}$ & 0 & 0 & 1.000 & 1.000 & 0 & 1.000 & 0.7143 & 1.000 & 0 & 0\\
$\alpha_{i3}$ & 0 &	0 & 0 & 0 & 1.000 & 0 & 0.2857 & 0 & 1.000 & 1.000\\
$b_i$ & 0 & 0 & 0 & 0 & 0 & 0 & 0.3773 & 0 & 0 & 0\\
$\xi_i$ & 0 & 0 & 0 & 0 & 0 & 0 & 0.5446 & 0 & 0 & 0\\
\hline
Vertex & GEHAM & ASARO & UHETO & SEUVE & NAGAD & GAMA\\
\hline
$\alpha_{i1}$ & 0 & 0 & 0 & 0 & 1.000 & 1.000\\
$\alpha_{i2}$ & 1.000 & 1.000 & 0 & 0 & 0 & 0\\
$\alpha_{i3}$ & 0 & 0 & 1.000 & 1.000 & 0 & 0\\
$b_i$ & 0 & 0 & 0 & 0 & 0 & 0\\
$\xi_i$ & 0 & 0 & 0 & 0 & 0 & 0\\ \hline
\hline
\end{tabular}
\label{tab2}
\end{table*}

Applying our model to this signed network, we find that the MDL achieves its minima when $c = 2$,
as shown in Fig.~\ref{fig8}(a), indicating that there are exactly two communities in the network.
Fig.~\ref{fig5}(a) gives the partition obtained by our method, which divides the network into two
groups of equal size and produces a completely consistent split with the true communities in the
network. As expected, vertices within the same community are mostly connected by positive links
while vertices from different communities are mainly connected by negative links. We shade each
vertex proportional to the parameters $\{\alpha_{ir}\}_{r=1}^c$, the magnitude of which supplies us
with the probabilities of each vertex belonging to different groups.\footnote{This network as well
as the Gahuku-Gama Subtribes network are both undirected graph, and therefore the parameter
$\alpha$ is identical to $\beta$, and $\theta$ is identical to $\phi$.} From Table~\ref{tab1},
we see that all the vertices can be exclusively separated into two communities, expect for the
vertex \lq\lq SNS\rq\rq~which belongs to the circle group with probability 0.0186 and to
the square group with probability 0.9814. In other words, the two communities overlap with each
other at this vertex, resulting in its high bridgeness of 0.0372 and group entropy of 0.1334. This is
validated by the observation that the vertex has two negative links with vertices \lq\lq
ZS-ESS\rq\rq~and \lq\lq DS\rq\rq~in the same community. We also visualize the learned parameters
$\omega^+$ and $\omega^-$ in Fig.~\ref{fig5}(b), which indeed provide a coarse-grained description
of the signed network and reveal that this network actually has two communities.

\begin{figure}[!t]
\centering
\includegraphics[width = 0.45\textwidth]{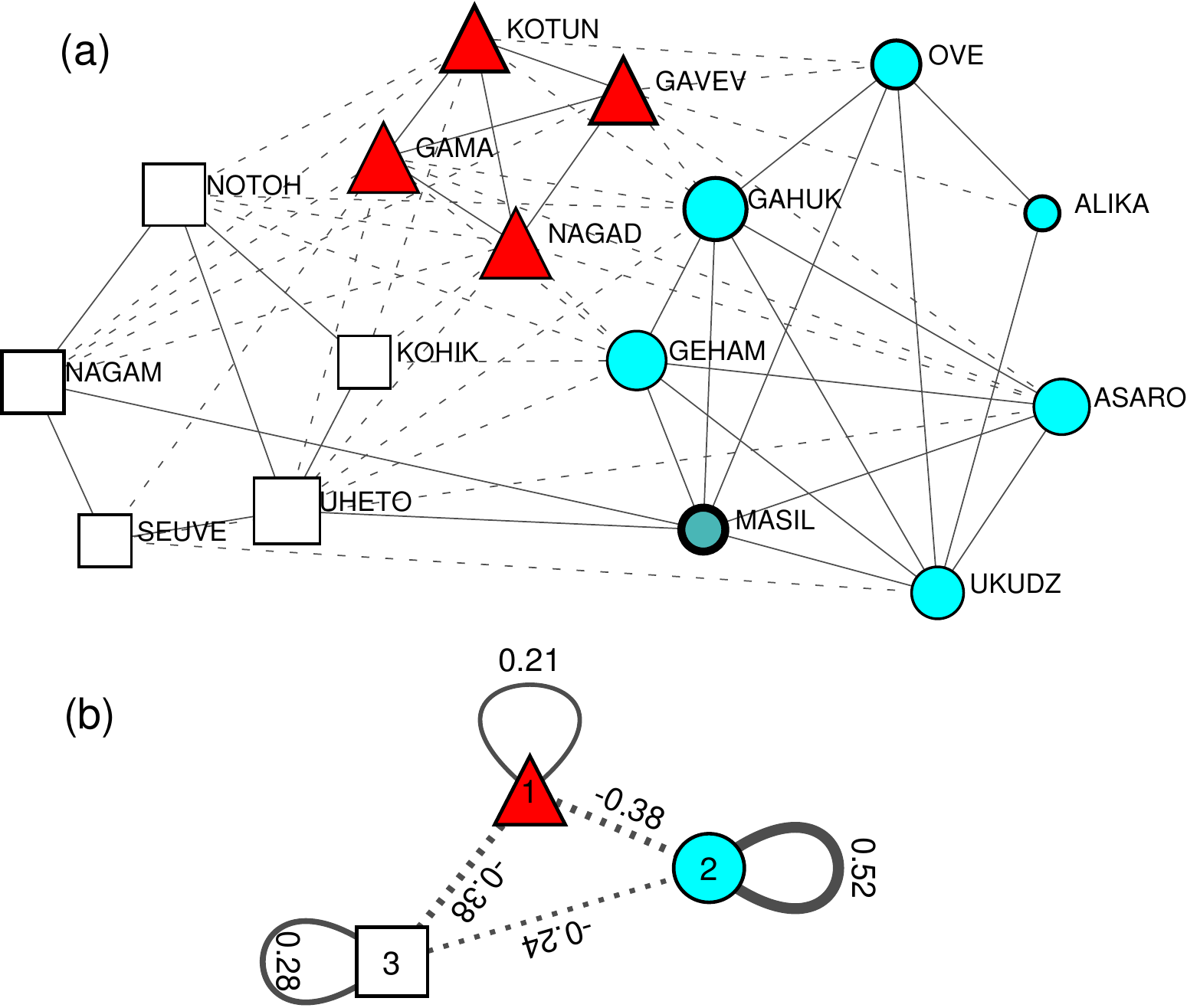}
\caption{(Color online) Exploratory analysis of the Gahuku-Gama Subtribes network~\cite{Read54}. The solid edges denote the positive links and the dashed edges represent negative links. The true
community structure in this network is represented by three different shapes while the inferred
groups are denoted by different colors. The sizes of the vertices are proportional
to the parameters $\theta$.}
\label{fig6}
\end{figure}

\begin{figure*}[!t]
\centering
\includegraphics[width = 0.9\textwidth]{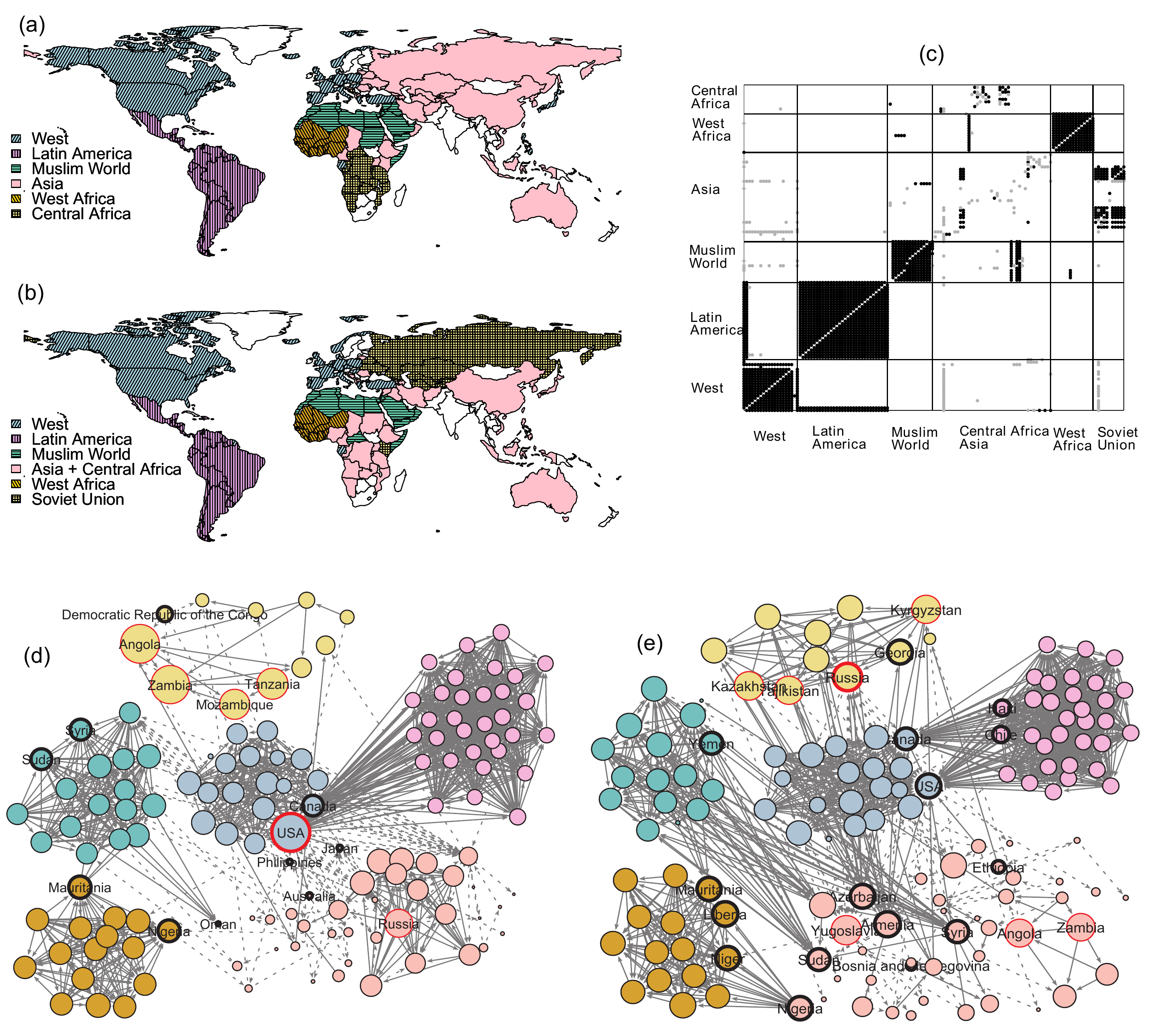}
\caption{(Color online) Exploratory analysis of the international conflict and alliance network~\cite{VJ09}. Maps of the groups found using the SSBM model from the perspective of outgoing edges (a) and incoming edges (b). (c) The rearrange adjacency matrix, in which the black dots denote positive links and the gray dots represent negative edges,  respectively. Six groups are separated by black solid lines. The partition of this network obtained by the SSBM model from the perspective of outgoing edges (d) and incoming edges (e), where the solid edges denote the positive links and the dashed edges represent negative links. The sizes of vertices are respectively proportional to their centrality degree $\theta$ and $\phi$. The red bold border vertices have the large centrality degrees while the black bold border vertices have the large values of bridgeness and group entropy.}
\label{fig7}
\end{figure*}

The second network is the Gahuku-Gama Subtribes network, which was created
based on Read's study on the cultures of Eastern Central Highlands of New
Guinea~\cite{Read54}. This network describes the political alliance and enmities among the
16 Gahuku-Gama subtribes, which were distributed in a particular area and were engaged
in warfare with one another in 1954. The positive and negative links of the
network correspond to political arrangements with positive and negative ties, respectively.
Fig.~\ref{fig8}(b) tells us that this signed network consists of three groups because
the MDL of the SSBM model is minimum when $c = 3$. The three groups categorized by our
model are given in Fig.~\ref{fig6}(a), and they match perfectly with the true
communities in the signed network. As shown in Table~\ref{tab2},
the vertex \lq\lq MASIL\rq\rq~participates in the circle group with probability 0.7143 and
in the square group with probability 0.2857. As a result, it has a large value of
bridgeness 0.3773 and group entropy 0.5446.  This implies that these two groups overlap
with each other at this vertex, which is approved by the fact that the vertex \lq\lq MASIL\rq\rq
has two positive links connected to \lq\lq NAGAM\rq\rq~and \lq\lq UHETO\rq\rq, respectively. The learned parameters $\omega^+$ and $\omega^-$ supply us with a thumbnail of the signed network again in Fig.~\ref{fig6}(b).

Finally we test our model on the network of international relation taken
from the Correlates of War data set over the period 1993---2001~\cite{VJ09}.
In this network, positive links represent military alliances and negative links denote military
disputes. The disputes are associated with three hostility levels, from
\lq\lq no militarized action\rq\rq~to \lq\lq interstate war\rq\rq.
For each pair of countries, we chose the mean level of hostility
between them over the given time interval as the weight of their negative link.
The positive links denote the alliances: 1 for entente, 2 for non-aggression pact and
3 for defence pact. Finally, we normalized both the negative links and positive links into
the interval [0, 1] and the final weight of the link among each pair of countries is
the remainder of the weight of the normalized positive links subtracting the weight
of the normalized negative links. The obtained network contains a giant component
consisting of 161 vertices (countries) and 2517 links (conflicts or alliances). Here, we
only investigate the structure of the giant component.

The structure of this network has been investigated in several existing
studies. These studies indicated that there are six main power blocs,
each consisting of a set of countries with similar actions of alliances or
disputes. In Ref.~\cite{VJ09}, the authors labeled these power blocs as (i) The West, (ii) Latin
America, (iii) Muslim World, (iv) Asia, (v) West Africa, and (vi) Central Africa.
Applying the SSBM model to this network, we find that the MDL arrives its minimum when $c = 6$,
as illustrated in Fig.~\ref{fig8}(c). By partitioning the network into six groups, we summarize
the results in Fig.~\ref{fig7}. From the rearranged adjacency matrix [Fig.~\ref{fig7}(c)],
we can conclude that the first, second, third and fifth groups, from bottom left to top right,
distinctly belong to the community structure, while the sixth group can be viewed as
the disassortative structure. However, the fourth group cannot be simply categorized as either
community structure or disassortative structure. In agreement with the assumption
of the SSBM model,  vertices in the six groups exhibit the similar connection profiles, although the miscellaneous structural patterns coexist in this network.

From the perspective of the outgoing edges, we obtain a split of the
network that is similar to the one got in Ref.~\cite{VJ09}, as shown in Fig.~\ref{fig7}(a). However, several
notable difference exists between the two results. Specifically, \lq\lq Pakistan\rq\rq~is grouped
with the West and \lq\lq South Korea\rq\rq~is grouped with the Muslim World in Ref.~\cite{VJ09}.
These false categorizations can be correctly amended, which is consistent with the
configuration depicted in Huntington's renowned book
\textit{The Clash of Civilizations}~\cite{Huntington96}. In addition, we categorized \lq\lq
Australia\rq\rq, which
is grouped with West in Ref.~\cite{VJ09}, into the group Asia for understandable reasons. Fig.~\ref{fig7}(b) gives a quite different structure of this network from the
perspective of incoming edges. Three groups, namely the West,
Latin America and Muslim World, stay almost the same. But \lq\lq Russia\rq\rq, together
with some countries of the former Soviet Union, are isolated from the Asia group and
form another independent power bloc. Meanwhile, the remaining countries in Asia group join with the West Africa countries to constitute a bigger cluster.
It is not difficult to see that all the changes appear to be in accordance with the
history and evolution of the international relations.

Recall that the parameters $\theta$ and $\phi$ provide us
with the centrality degrees of each vertex in its corresponding group
from the perspective of outgoing edges and incoming edges, respectively.
In other words, the parameters measure the importance of each vertex in its group.
For a better visualization, the sizes of vertices in Fig.~\ref{fig7}(d) and (e)
are proportional to the magnitude of the scalars $\theta$ and $\phi$.
Coincidentally, we discover that the big vertices, marked by the red bold border,
usually stand for the dominant countries in their corresponding groups. For example, the largest
vertex of the West is \lq\lq USA\rq\rq~in Fig.~\ref{fig7}(d). In fact,
this state often serves as a leader in its power bloc. A similar
interpretation can be given for the vertex \lq\lq Russia\rq\rq~in Asia group.
We further check the bridgeness and group entropy for each vertex in the network (data not shown),
and we mark the vertices, which have large values of these two measures, with the black bold
border. As anticipated, these kinds of vertices are
particularly prone to reside on the boundaries of different groups.
That is to say, the vertices that are very difficult to divide into one group build a fuzzy watershed of the overlapping structures. In Fig.~\ref{fig7}(b), three vertices
\lq\lq Janpan\rq\rq~, \lq\lq Philippines\rq\rq~and \lq\lq Australia\rq\rq, with high values
of bridgeness and group entropy,
play a transitional role between the West and Asia groups. In reality,
the above-mentioned Asian counties frequently collaborated with
the counterparts in West group in many areas, from economics to military.

\section{Conclusions}
\label{sec5}

We propose an extension of the stochastic block model to study
the mesoscopic structural patterns in signed networks.
Without prior knowledge what specific structure exists, our model can not only accurately detect broad types of intrinsic structures, but also can directly learn their types from the network data. Experiments on a number of synthetic and real world networks demonstrate
that our model outperforms the state-of-the-art approaches
at extracting various structural features in a given network.
Due to the flexibility inherited from the stochastic model,
our method is an effective way to reveal the global organization
of the networks in terms of the structural regularities, which further
helps us understand the relationship between networks' structure
and function. As future work, we will generalize
our model by releasing the requirement that the block
matrices are square matrices and investigate the possible
applications of the more flexible models.

\begin{acknowledgments}
The author would like to thank Vincent A. Traag for providing the international conflict and alliance network used in this paper. The author is also grateful to the anonymous reviewers for their valuable suggestions, which were very helpful for improving the manuscript.
\end{acknowledgments}

\end{document}